\documentclass[12pt,a4paper]{cibb}

\makeatletter
\providecommand{\@ordinalM}[2]{#1}
\makeatother

\usepackage{subfigure,graphicx}
\usepackage{amsmath,amsfonts,latexsym,amssymb,euscript,xr}
\usepackage{booktabs}
\usepackage[nodayofweek]{datetime}
\usepackage{hyperref}
\usepackage{fmtcount}
\usepackage[english]{datenumber}
\usepackage[absolute]{textpos}
\usepackage{natbib}

\usepackage{tikz}

\usepackage{algorithmicx}

\usepackage[table]{xcolor}
\usepackage{color,colortbl,tabularx}

\usepackage[english]{babel}
\usepackage[protrusion=true,expansion=true]{microtype}
\usepackage{amsmath,amsfonts,amsthm}
\usepackage{pifont}

\newcommand{\code}[1]{\texttt{\small #1}}

\definecolor{LightBlue}{rgb}{0.88,0.9,0.9}

\title{\Large $\ $\\ \bf A Block Decomposed QUBO Workflow for Chromosome-Y Phylogeny Reconstruction}

\author{\large Giuliana Siddi Moreau$^{*,1}$, Riccardo Berutti$^{1}$, Manuela Profir$^{1}$, Lorenzo Pisani$^{1}$, Maria Laura Clemente$^{1}$, and Lidia Leoni$^{1}$}
\address{ \footnotesize $\ $\\$^1$ CRS4 - Centro di Ricerca, Sviluppo e Studi Superiori in Sardegna, Pula, Italy. \\
\bigskip
ORCID codes: GSM 0000-0002-0945-6915; RB 0000-0003-1862-3700; MP 0000-0001-6996-6071; LP 0000-0002-2956-5668; MLC 0000-0002-5952-2810; LL 0000-0001-5262-3275.
\bigskip
\newline
$^*$corresponding author: julie@crs4.it
\bigskip
\newline
}
\abstract{\small QUBO, ADMM decomposition , Counter-diabatic quantum
optimisation , Phylogenetics, Y-chromosome. \normalsize
\\[17pt]
{\bf Abstract} This paper sets out a computational workflow that reconstructs the phylogeny of human Y-chromosome populations from a Variant Call Format (VCF) file of biallelic Single Nucleotide Polymorphisms (SNP). The two classical phylogenetic decisions – topology selection and root placement – are cast as Quadratic Unconstrained Binary Optimisation (QUBO) problems.
The workflow combines two QUBO formulations with an Alternating Direction Method of Multipliers (ADMM) decomposition strategy and a plug-in Digitized Counter-Diabatic Quantum Optimization (DCQO) solver, enabling large phylogenetic optimization problems to be executed across multiple classical or quantum computing resources.
  The DCQO gate-based optimizer drives each ADMM-block QUBO with short-depth counter-diabatic circuits in the impulse regime, without the necessity for an outer variational loop.
The combination of ADMM decomposition and the DCQO block solver facilitates problem sizes that surpass the qubit budget of any individual digital quantum processing unit call, while maintaining the integrity of the original objective.
The workflow extracts genotype information from VCF files, reconstructs topology and rooting through two QUBO formulations, and annotates the resulting tree using PhyloTree.
 
The resultant data set comprises a Nexus-annotated rooted tree, in addition to  diagnostic figures.  The workflow demonstrates the potential of modest-scale QUBO formulations combined with ADMM decomposition to serve as a scalable alternative to greedy heuristics in the domain of population genomics.

}

\footnotetext{\small{Article version: \datedate $\;$ h\currenttime  $\;$ CET}}

\begin{document}

\thispagestyle{myheadings}
\pagestyle{myheadings}
\markright{\tt Proceedings of CIBB 2026}

\section{Introduction}
\label{sec:SCIENTIFIC-BACKGROUND}
Reconstructing evolutionary relationships among biological entities is one of the central tasks in computational biology.
Given $n$ taxa, the number of possible unrooted binary tree topologies is $(2n-5)!!$,
 which grows super-exponentially and renders exhaustive search infeasible for even moderate values of $n$.
This combinatorial explosion has motivated decades of heuristic development, from distance-based methods such as Neighbor-Joining  to optimality-criterion approaches including maximum parsimony  and maximum likelihood.

Quantum computing offers a fundamentally new paradigm for attacking combinatorial optimization problems.
In particular, quantum annealing \citep{farhi2000} and the Quantum Approximate Optimization Algorithm (QAOA) \citep{farhi2014} have shown promise on NP-hard problems when formulated as Quadratic Unconstrained Binary Optimization (QUBO) or Ising models.
Recent work has begun to connect these quantum approaches with phylogenetics: Onodera et al.\ \citep{onodera2023} demonstrated phylogenetic tree reconstruction via normalized graph-cut on Fujitsu's Digital Annealer; Dinneen et al.\ \citep{DINNEEN202360} formulated the Tree Containment problem as QUBO; Bach et al.\ \citep{bach2024} encoded maximum parsimony as a Steiner tree QUBO; and Zhang et al.\ \citep{zhang2026} proposed an efficient branch-based model compatible with both classical and quantum solvers.

Despite these individual advances, no unified framework exists that (i) implements multiple QUBO formulations for phylogenetics within a common architecture, (ii) provides device-agnostic routing across quantum annealing and gate-based backends.

The main contributions of this work are fourfold. First, we formulate both topology selection and root placement as QUBO optimization problems within a unified phylogenetic workflow. Second, we introduce an ADMM-based decomposition strategy that enables large QUBO instances to be partitioned into overlapping subproblems. Third, we integrate a DCQO solver as a plug-in optimizer for each ADMM block. Finally, we demonstrate the complete workflow on chromosome-Y phylogeny reconstruction from Variant Call Format (VCF) data.

\section{Data and Methods}
\label{sec:DATA-AND-METHODS}
Fig. \ref{fig:flow} summarizes the overall workflow. Starting from a VCF file, genotype information is converted into a distance matrix, a Neighbor-Joining tree is generated to identify candidate splits, two QUBO formulations optimize topology and rooting, and the resulting tree is finally annotated using PhyloTree.

The reconstruction of the topology and root of a population tree from genome-scale variant data constitutes a two-stage decision problem.  The initial step involves the selection of a set of mutually compatible bipartitions of the leaf set, with the objective of providing a comprehensive explanation for the observed distance matrix. Subsequently, the second step entails the selection of the edge on which to insert the root.  It is evident that both of these processes are of a combinatorial nature, and both are routinely addressed by means of greedy heuristics (NJ, FastME, midpoint rooting). Furthermore, both of these processes are amenable to a natural QUBO encoding, which is suitable for simulated or quantum annealing.

This contribution presents a self-contained Python workflow
 that implements the 
twin-QUBO formulation for human chrY data. Tree leaves are individual VCF samples (genotypes parsed from the \code{FORMAT}/\code{GT} fields in the sample columns), and branches are coloured by their dominant
PhyloTree-Y haplogroup, so that the QUBO-reconstructed clades can be
read off the tree at a glance.  The workflow depends
only on widely-used libraries ({\it cyvcf2}, {\it scikit-bio}, {\it dimod},
{\it dwave-neal}), and produces a NEXUS file compatible with FigTree, iTOL
and BEAST-derived viewers.  Branches are annotated with the marker names and
clades of \emph{PhyloTree Y} \cite{vanOven2014}, retrieved at runtime from
{\it phylotree.org}.

A central contribution of this work is a strategy for executing large phylogenetic QUBOs beyond the capacity of a single optimizer. The proposed Alternating Direction Method of Multipliers (ADMM) decomposition partitions the global problem into overlapping blocks that can be solved independently using simulated annealing or digital quantum optimization before being reconciled through consensus updates ~\cite{boyd2011}. 

Crucially, the per-block solver is itself pluggable: in addition to
classical simulated annealing we provide a pure DCQO gate-based solver implemented 
following Dalal et al.~\cite{dalal2024} and building on the theoretical
foundations of~\cite{hegade2022}.  The block QUBO is converted
to its Ising form and a
short-depth circuit is synthesised from the closed-form first-order counter-diabatic
coefficient $\alpha_1(t)$ (Eq.~(6) of \cite{dalal2024}) and executed in
the impulse regime: no outer variational loop is needed.
The workflow is hence forward-compatible with near-term
QPU hardware without requiring the global QUBO to fit in a single
embedding.

\section{Pipeline Overview}

\begin{figure}[ht]
\centering
\includegraphics[width=0.9\linewidth]{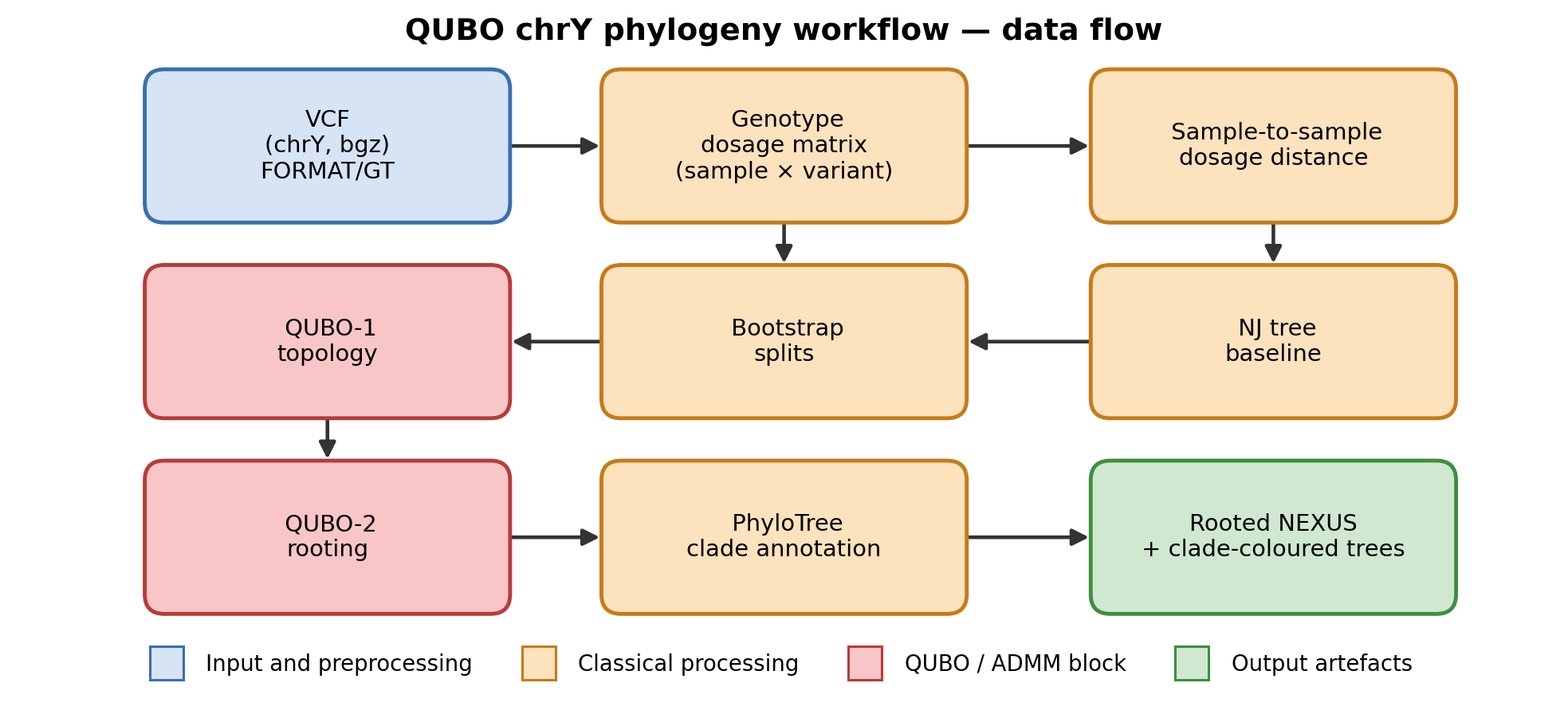}
\caption{Data-flow of the workflow.  Blue = input and preprocessing (the VCF read via
         FORMAT/GT columns), orange = classical processing (genotype
         dosage matrix, sample-to-sample distance, NJ baseline,
         bootstrap, PhyloTree-Y clade annotation), red = QUBO optimization blocks (topology optimization, rooting optimization, and the shared ADMM/DCQO solver), green = output artefacts (rooted NEXUS plus the
         clade-coloured tree plots).}
\label{fig:flow}
\end{figure}

The workflow is organised in steps that follow a strict data-flow order.  Figure~\ref{fig:flow} summarises
the eight functional stages.

During VCF ingestion variants are loaded directly from the per-sample columns that follow the
\code{FORMAT} field of the VCF;  Each
sample's diploid (or haploid) genotype \code{GT} is converted to an
\emph{ALT-dosage} scalar
\begin{equation}
   d_{n,v} \;=\;
   \begin{cases}
      0     & \text{if the called allele is REF (0/0, 0)} \\
      1     & \text{if the called allele is ALT (1/1, 1)} \\
      \text{NaN} & \text{if fully missing (./., .)},
   \end{cases}
\end{equation}
producing an $N\!\times\!M$ matrix $D\in\{0,1\}^{N\times M}\cup\{\text{NaN}\}$
of $N$ samples by $M$ variants.  Records are retained only when they pass
the \code{FILTER} column,  present a per-site missing
rate $\leq$ \code{max-missing} (default $0.8$) and a carrier rate
$\geq$ \code{min-af} (default $0.001$, here meaning the fraction of called
samples with $d>0$).  Remaining NaNs are imputed to $0$.   Female subjects, even if not identified in advance, are removed automatically on the basis of their lack of informativeness.
Tree leaves are the $N$ samples, labelled by the $1$-based
position of their column after \code{FORMAT}.

Next, a sample-to-sample distance matrix is computed directly from the dosage matrix
as the mean absolute dosage difference across variants,
\begin{equation}
   d_{ij} \;=\;
   \frac{1}{M}\sum_{v=1}^{M} \bigl|\,d_{i,v}-d_{j,v}\,\bigr|,
   \qquad i,j\in\{1,\dots,N\},
\end{equation}
yielding a symmetric, non-negative $N\!\times\!N$ matrix that drives the following two phases.

A Neighbor-Joining tree is then constructed in order to define its internal edges
as  the set of splits used both as a validation oracle and as one of the
ingredients of the bootstrap procedure.

Bootstrap resampling (200 replicates over colunms of  $D$) identifies candidate splits.
 For
each replicate a fresh sample-to-sample distance matrix and NJ tree are
computed and every non-trivial bipartition of the $N$ leaves is hashed in
canonical form.  Splits with $\geq 50\%$ support are retained and the top
$K=30$ feed the QUBO step.

During the split support and compatibility phases, each candidate split $S=(A,B)$ receives a continuous support score
\begin{equation}
  \sigma(S) \;=\; \bar d_{AB} - \tfrac{1}{2}(\bar d_{AA}+\bar d_{BB}),
\end{equation}
i.e.~the gap between mean between-group and within-group distance.  Two
splits $(A,B)$ and $(C,D)$ are \emph{compatible} when one of the four
pairwise intersections $A\!\cap\!C$, $A\!\cap\!D$, $B\!\cap\!C$,
$B\!\cap\!D$ is empty, i.e. the standard four-points condition for trees is imposed.

QUBO-1 selects a subset of mutually compatible candidate splits. The objective rewards highly supported splits while penalizing incompatible combinations and enforcing the expected number of internal branches.

To build  QUBO-1, that governs the topology selection, the binary variable $x_i\in\{0,1\}$ is associated with every candidate split
$S_i$ in the following fashion:
\begin{equation}
  H_{\text{top}} \;=\;
     -\!\!\sum_{i} \hat\sigma_i\,x_i
     \;+\;\lambda \!\!\!\sum_{(i,j)\,\text{incompatible}}\!\!\! x_i x_j
     \;+\;\mu\!\left(\sum_{i} x_i - k\right)^{\!2},
\label{eq:qubo1}
\end{equation}
with $\hat\sigma_i$ the support normalised to $[-1,1]$, $\lambda=5$ the
incompatibility penalty, $\mu=2.5$ the budget penalty, and $k=N-3$ the
expected number of internal splits for a binary unrooted tree on
$N$ leaves (samples). 

Once the topology has been selected, the remaining task is to determine the most plausible root location. This is formulated as a second QUBO. 
The unrooted tree is then converted to a weighted graph.  For
every edge $e$ a candidate root is inserted at its midpoint and the variance
of the resulting root-to-tip distances is recorded as
$c_e=\operatorname{Var}_{\ell}[d(\text{root},\ell)]$ -- the deviation from a
molecular clock.  A binary variable $r_e$ per edge enters the QUBO-2
\begin{equation}
  H_{\text{root}} \;=\;
     \sum_e \hat c_e\,r_e
     \;+\;\eta\!\left(\sum_e r_e - 1\right)^{\!2}, \label{eq:qubo2}
\end{equation}
with $\hat c_e$ the normalised cost and $\eta=10$ enforcing the
\emph{exactly one root} constraint.

Branch-specific mutations are inferred by mapping ALT-dosage variants to the most recent common ancestor of their carrier set; the resulting list of mutations is
attached to the corresponding branch and truncated to
a threshold value on the number of entries.

Finally, branch annotations are enriched using PhyloTree and
their PhyloTree names propagated to the final NEXUS file.

Eqs.~\ref{eq:qubo1} and \ref{eq:qubo2} can be solved with
a simulated annealing sampler, a quantum annealer  or with ADMM block-decomposed digital quantum optimization solver on multiple nodes.

\section{ADMM decomposition \label{sec:admm}}
As the number of candidate splits increases, the resulting QUBO rapidly exceeds the size that can be handled efficiently by a single annealer or quantum processor. To overcome this limitation we partition the optimization problem into overlapping subproblems that are coordinated using ADMM.

The candidate split selection in Eq.~\ref{eq:qubo1} and the rooting Hamiltonian in Eq.~\ref{eq:qubo2} remain tractable for
the chrY application reported here, but both are written so that a much
larger candidate pool (e.g.~all $2^{p-1}\!-\!1$ bipartitions, or per-branch
length quantisations) can be plugged in unchanged.  

The key idea is to replace one large optimization problem by several smaller overlapping optimization problems whose solutions are iteratively reconciled through consensus variables.  To make those regimes
practical, we implemented an ADMM-based decomposition solver and a user-supplied partition
$\mathcal{V}=\{V_1,\dots,V_K\}$ of its variables, with overlapping
\emph{coupling sets} $C_{kl}=V_k\cap V_l$.

Let $\mathbf{x}=(x_v)_{v\in V}\in\{0,1\}^{|V|}$ be the binary vector of the
full QUBO $H(\mathbf{x})=\mathbf{x}^{\!\top}\!Q\,\mathbf{x}$.  Introducing
block-local copies $\mathbf{z}^{(k)}$ for each $V_k$ and consensus variables
$\boldsymbol{\bar z}$ on the overlaps, we minimise the augmented Lagrangian
\begin{equation}
  \mathcal{L}_\rho \;=\;
    \sum_{k=1}^{K} H_k(\mathbf{z}^{(k)})
    \;+\; \sum_{k}\sum_{v\in C_k}\!
            \bigl[\lambda^{(k)}_v (z^{(k)}_v - \bar z_v)
                 + \tfrac{\rho}{2}(z^{(k)}_v - \bar z_v)^{2}\bigr],
\label{eq:admm}
\end{equation}
where $H_k$ is the restriction of $H$ to $V_k$ (with cross-block couplings
absorbed into linear biases via $\bar z$), $\lambda^{(k)}_v$ are dual
multipliers, and $\rho>0$ is the penalty weight.  Because each
$z^{(k)}_v\in\{0,1\}$, the quadratic penalty
$\tfrac{\rho}{2}(z^{(k)}_v - \bar z_v)^{2}$ remains quadratic-binary after
expansion and is folded back into a local QUBO solver call.  

ADMM alternates between three steps: (i) solving each local QUBO independently, (ii) updating the consensus variables shared among blocks, and (iii) updating the dual variables until the primal and dual residuals satisfy the stopping criteria.

%
Convergence is achieved when the primal residual
$\|\mathbf{z}^{(k)}-\boldsymbol{\bar z}\|_2$ and the dual residual
$\rho\|\boldsymbol{\bar z}^{(t)}-\boldsymbol{\bar z}^{(t-1)}\|_2$ fall below
user-set thresholds, or after a maximum number of rounds. 
Empirically the topology QUBO converges in 6--12 ADMM rounds
with two-block partitions for $|\mathcal{S}|\!\leq\!60$, and the final
energy matches the monolithic simulated annealing solution to within numerical precision on
the reference dataset.  

\section{Per-block optimizer: digitized counter-diabatic QUBO solver.}
Step~(i) of the ADMM iteration calls a quantum computing solver that minimises
the local penalised QUBO.  We implement this solver as a  digitized counter-diabatic quantum
optimisation (DCQO) routine that follows the algorithmic recipe of
Dalal et al.~\cite{dalal2024}. The corresponding counter-diabatic circuit is generated analytically in the impulse regime, avoiding an outer variational optimization loop.

The local block QUBO, containing the ADMM augmented-Lagrangian penalty
$\tfrac{\rho}{2}\!\sum_{v\in C_k}(z^{(k)}_v-\bar z_v+\lambda^{(k)}_v/\rho)^{2}$
already folded in, is first cast in Ising form via
$x_v=\tfrac{1}{2}(1-z_v)$, yielding
$H_{\!p}=\sum_{i\in V_k} h_i Z_i + \sum_{i<j\in V_k} J_{ij}\,Z_i Z_j$
with $J_{ij}=Q_{ij}/2$, $h_i=-Q_{ii}/2-\tfrac{1}{2}\sum_{j\neq i}Q_{ij}$.
The circuit starts from the uniform superposition
$|+\rangle^{\otimes|V_k|}$ (the ground state of the mixer
$H_{\!m}=-\sum_{i\in V_k} X_i$ ) and a time-dependent
counter-diabatic generator
$H_{\!\text{CD}}(t)= \alpha_1(t)\,\bigl(\sum_i h_i Y_i +
                                       \sum_{i<j} J_{ij}\,(Y_iZ_j+Z_iY_j)\bigr)$
is Trotterised into $N$ steps under the smooth schedule
\begin{equation}
   \lambda(t) \;=\; \sin^{2}\!\Bigl(\tfrac{\pi}{2}\,
                     \sin^{2}\!\bigl(\tfrac{\pi t}{2T}\bigr)\Bigr),
   \qquad t\in[0,T],
\end{equation}
whose first and second derivatives vanish at the endpoints.  At each
Trotter step $m$ the analytic first-order CD coefficient
$\alpha_1(\lambda(t_m))$ is computed in closed form from $\{h_i,J_{ij}\}$
(Eqs.~(5)--(7) of \cite{dalal2024}) and combined with the per-step
prefactor
$\theta_m=\tfrac{\pi}{2N}\sin\!\bigl(\tfrac{\pi m}{N}\bigr)
                     \sin\!\bigl(\pi\sin^{2}\!\tfrac{\pi m}{2N}\bigr)$.
Each Trotter step therefore consists of single-qubit $R_y$ rotations of
angle $-2\theta_m\alpha_1 h_i$ and two-qubit $\exp(-i\phi\,Y_iZ_j)$,
$\exp(-i\phi\,Z_iY_j)$ rotations of angle $\phi=-2\theta_m\alpha_1 J_{ij}$,
each decomposed into a $\{S^{\dagger}, \text{CNOT}, R_y, \text{CNOT}, S\}$
sequence.  Rotations whose angle falls below
a preset gate cutoff threshold (default $0.1$\,rad) are dropped so as to
keep the realised circuit shallow.

The optimizer runs in the pure-DCQO, impulse-regime mode: the circuit is
constructed once for $N$ Trotter steps from the analytic CD coefficients
and sampled in a single shot batch.

 Since each ADMM block contains approximately
$|V_k|\!\lesssim\!12$--$15$ qubits, the resulting
circuits are shallow and stay comfortably within the two-qubit-gate
coherence budget of present-day superconducting and trapped-ion
processors.  Best-of-shots bitstrings are returned to the ADMM consensus
step exactly as if they had been produced by simulated annealing, making the two solvers
interchangeable from the driver's point of view.

\section{Results}
The workflow was benchmarked on 78 populations from the gnomAD v3.1.2 HGDP+1KG chromosome-Y callset, retaining 128,316 biallelic SNPs across the panel after per-population AF-field parsing (sex-stratified and aggregate strata removed, non-informative populations dropped), and successfully produced a rooted phylogenetic tree. Both QUBO instances were solved with the ADMM decomposition driver dispatching each block to a pure-DCQO sub-solver (2000 shots, 4 Trotter steps; blocks capped at 18 qubits). The bootstrap stage (200 replicates) returned $|\mathcal{S}| = 58$
 candidate splits at $\geq 50\%$  support; using the 30 highest-support splits, the topology QUBO comprised 30 binary variables (435 quadratic couplings) and reached its optimum (energy $5050.80$) in $95.1\;\mathrm{s}$
 of wall time, with all $30$ selected splits recovered in the Neighbor-Joining reference tree. The rooting QUBO comprised $153$ binary variables, one per candidate insertion edge on the unrooted binary tree of 78 leaves ($11,628$ quadratic couplings), which the ADMM driver decomposed into 16 blocks and solved over 9 outer iterations in $652.5\,\mathrm{s}$
of wall time, returning a best solution of energy $-9.61$ and placing the root on branch  at root-to-tip variance $\sigma^2 = 6.26\times10^{-3}$.

\begin{figure}
    \centering
    \includegraphics[width=0.80\linewidth]{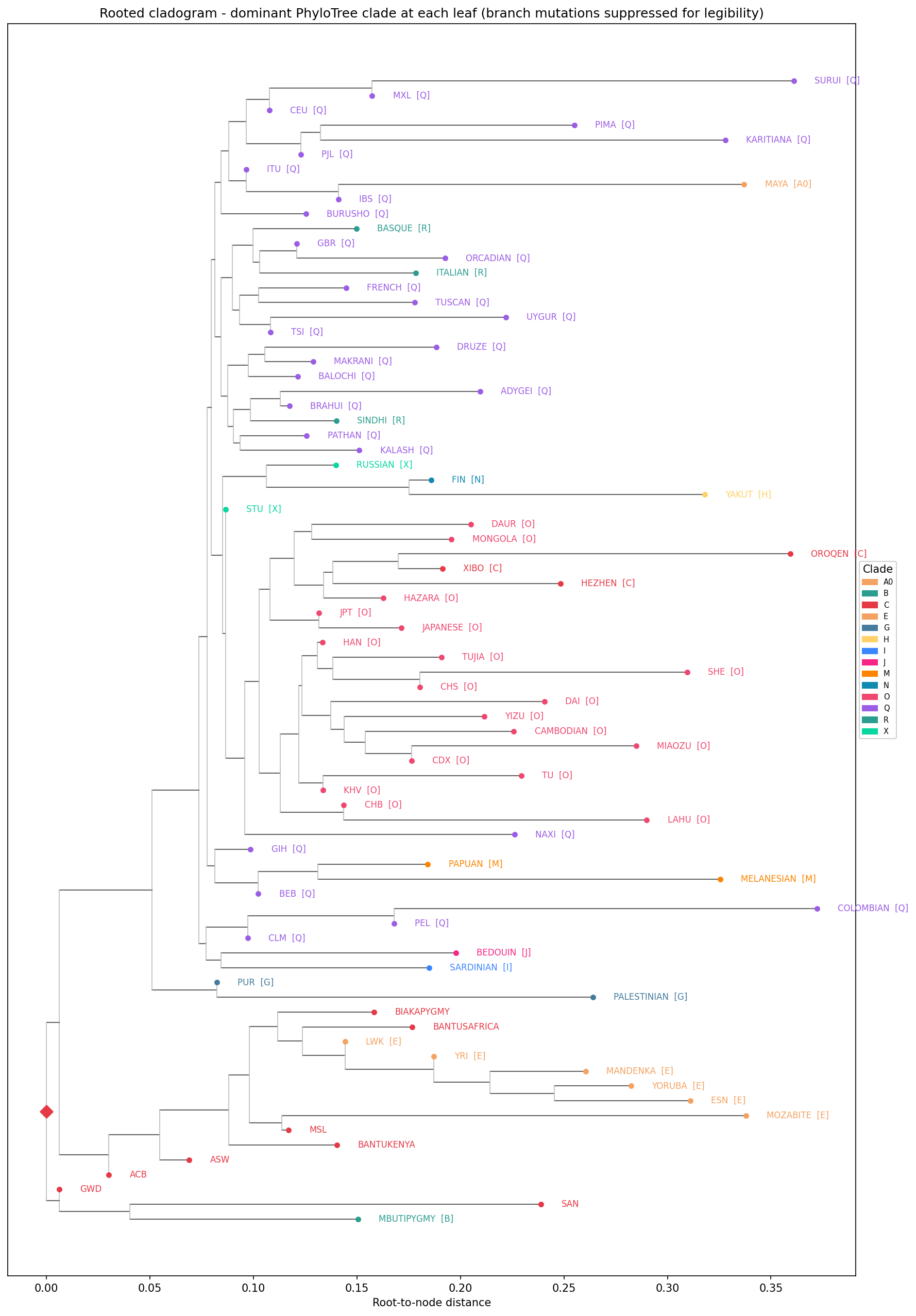}
    \caption{Rectangular phylogram of 78 male chromosome-Y samples reconstructed from a 150-sample gnomAD GRCh38 callset (chrY VCF), after removing 72 uninformative samples with no ALT call across the retained variants. The tree topology is obtained from the QUBO topology solve (QUBO-1), with branches placed on the rooting edge selected by the QUBO rooting solve (QUBO-2); both QUBOs are dispatched to the ADMM service with DCQO block solvers running on noise-free Cirq simulator.}. 
    \label{fig:phylogram}
\end{figure}
Fig. \ref{fig:phylogram} shows the rectangular phylogram obtained in the run. The results are comparable to the reference run obtained via simulated annealing. 
All 30 QUBO-selected splits were consistent with Neighbor-Joining (precision 1.00, recall 0.40), and the inferred root recovered the minimum-variance branch exactly (rank 1/153), rooting the tree in deep African lineages as expected.

\section{Conclusion}
\label{sec:CONCLUSIONS}
Unlike previous quantum phylogenetic approaches that formulate isolated optimization problems, the proposed workflow combines scalable ADMM decomposition with a plug-in DCQO optimizer, enabling execution on heterogeneous quantum resources while preserving a unified optimization framework.
The presented workflow demonstrates that QUBO is a viable formalism for
small-to-medium phylogenetic decisions:
problem sizes fit comfortably on classical simulated annealing hardware and would map onto
near-term quantum annealers with minor edits.  The ADMM decomposition
solver (Section~\ref{sec:admm}) extends this reach further by replacing a
single monolithic QUBO call with a sequence of smaller block calls, each
solved by a digitized counter-diabatic circuit short enough to fit the
coherence budget of present-day gate-model QPUs. This decomposition allows each block to fit within the qubit capacity of current gate-based quantum processors while preserving the global optimization objective through consensus constraints.
Future work includes
(i) benchmarking the DCQO block optimizer on superconducting and trapped-ion
hardware against the simulated annealing baseline reported here and classical phylogenetic solvers,
(ii) extending the topology QUBO with branch-length variables, whose
multiplicative growth in the variable count is precisely the regime
ADMM\,+\,DCQO is designed for,
(iii) replacing the first-order counter-diabatic term with a variationally
optimised gauge potential to further shorten the per-block circuit, and
(iv) integrating a maximization step that re-estimates mutation probabilities from
the inferred tree to refine the support scores.

\section*{Conflict of interests}
\label{sec:CONFLICT-OF-INTERESTS}
The authors have no conflicts of interest to declare that are relevant to the content of this article.

\section*{Acknowledgments}
\label{sec:ACKNOWLEDGMENTS}
During the preparation of this work, the authors utilized Artificial Intelligence (AI) tools to assist with text drafting, language editing, and software development. Specifically, Claude 3.8 Opus was employed for drafting and to support the writing and optimization of the software code. ChatGPT was used for language editing and stylistic refinement. Following the use of these tools, the authors independently reviewed, verified, and edited all outputs to ensure scientific accuracy and integrity. The authors take full and sole responsibility for the final content, software code, and conclusions of this publication.

\section*{Funding}
\label{sec:FUNDING}
This work was carried out with the financial contribution of the Sardinia Regional Authorities.


\footnotesize
\bibliographystyle{unsrt}
\bibliography{bibliography_CIBB_file.bib} 
\normalsize

\end{document}